\documentclass[aps,nofootinbib]{revtex4}
\usepackage{amsmath}
\usepackage{amsfonts}
\usepackage{indentfirst}
\usepackage{graphicx}

\def\s{\textrm{s}}
\def\b{\textrm{b}}
\def\bs{\textrm{bs}}
\def\ss{\textrm{ss}}

\begin{document}

\title{Thermodynamics and Stability of Flat Anti-de Sitter Black Strings}
\author{Si Chen}
\email{sichen@phas.ubc.ca}
\author{Kristin Schleich}
\email{schleich@phas.ubc.ca}
\author{Donald M. Witt}
\email{donwitt@noether.physics.ubc.ca}
\affiliation{Department of Physics and Astronomy,
University of British Columbia,\\
6224 Agricultural Road,
Vancouver, BC V6T 1Z1, Canada\\ }
\date{\today}

\begin{abstract}
We examine the thermodynamics and stability of 5-dimensional flat anti-de Sitter (AdS)  black strings,  locally asymptotically anti-de Sitter spacetimes whose spatial sections are AdS black holes with Ricci flat horizons. We find that there is a phase transition for the flat AdS black string when the AdS soliton string is chosen as the thermal background. We find that this bulk phase transition corresponds to a 4-dimensional flat AdS black hole to AdS soliton phase transition on the boundary Karch-Randall branes. We compute the possibility of  a phase transition from a flat AdS black string to a 5-dimensional AdS soliton and show that, though possible for certain thin black strings, the transition to the AdS soliton string is preferred. In contrast to the case of the Schwarzschild-AdS black string, we find that  the specific heat of the flat AdS black string is always positive; hence it is thermodynamically stable. We show numerically that both the flat AdS black string and AdS soliton string are free of a Gregory-Laflamme instability for all values of the mass parameter. Therefore the Gubser-Mitra conjecture holds for these spacetimes.\end{abstract}
\pacs{11.25.Tq, 04.70.Dy, 04.50+h}
\maketitle

\section{Introduction}

The thermodynamics of asymptotically anti-de Sitter (AdS) spacetimes is interesting both in its own right and in its connection to the AdS/CFT correspondence.  As shown by Hawking and Page,  black holes in AdS spacetimes exhibit a first order phase transition to a thermal AdS background \cite{Hawking:1982dh}.  However, in marked contrast to the asymptotically flat case, the specific heat of sufficiently large black holes is positive in AdS spacetimes.  Therefore the canonical ensemble is well defined for these spacetimes without imposition of further conditions.  In the context of AdS/CFT, Witten showed that the Hawking-Page phase transition corresponds to a deconfinement-confinement phase transition in the large $N$ limit of supersymmetric Yang-Mills gauge theory on the conformal boundary \cite{Witten:1998zw}.  Thus the thermodynamics of asymptotically AdS spacetimes  also provides a useful probe in the study of  the AdS/CFT correspondence.

The study of the thermodynamics of black strings and black p-branes, higher dimensional spacetimes whose foliation yields a black hole on each spatial hypersurface, has also proven informative in the study of braneworlds.  In these scenarios, an alternative to more traditional Kaluza-Klein compactification, the 4-dimensional universe corresponds to a brane embedded in a higher dimensional anti-de Sitter spacetime.  The Randall-Sundrum model \cite{Randall:1999vf}, a particularly useful example of this
scenario, was extended by Karch and Randall to branes that are AdS slicings, corresponding to the case where the brane tension is slightly detuned from the critical Randall-Sundrum value \cite{Karch:2000ct}.
This extension provides an advantageous forum in which to study the physics of braneworld black holes as the thermodynamic stability of AdS black holes can be exploited. A first step, taken by Chamblin and Karch, utilized the Karch-Randall (KR) braneworld formulation to study the relation between bulk phase transitions of a black AdS string and those of the braneworld black hole \cite{Chamblin:2004vr}. They found an exact correspondence between the thermodynamics of the AdS black string in the bulk and that of the AdS black hole on the brane. The Hawking-Page phase transition on the brane, corresponding to a CFT phase transition on AdS$_{d-1} $ weakly coupled to gravity, is described by a phase transition between the black string and AdS$_{d}$ in the bulk. Furthermore, the temperature at which the specific heat of the black hole on the brane becomes negative, leading to black hole evaporation on the brane to a thermal AdS$_{d-1}$ state, qualitatively corresponds to the onset of the black AdS string instability for  Schwarzschild-AdS slicings found by Hirayama and Kang \cite{Hirayama:2001bi,Kang:2004ys}.   This analysis of the thermodynamics of black holes on asymptotically AdS branes provides a starting point for the further study of properties of black holes on the brane and their connection to the AdS/CFT correspondence, in particular the holographic conjecture for black holes on the brane \cite{Gregory:2008br}  (See \cite{Gregory:2008rf} for a recent review).

Given the usefulness of this model, it is natural to explore the connections between bulk and brane black hole phase transitions for other asymptotically AdS spacetimes. A natural starting point is see if this connection can be extended to AdS black hole solutions whose horizons have non-spherical topology.
The thermodynamics of  such spacetimes and various correspondences in a conformal field theory on the conformal boundary have been well studied. The pioneering work of Vanzo on the thermodynamics of AdS black holes with non-spherical horizons  \cite{Vanzo:1997gw} was extended and linked to the AdS/CFT correspondence in
\cite{Brill:1997mf,Birmingham:1998nr,Emparan:1999gf}.  Surya, Schleich and Witt found that Ricci flat AdS black holes exhibit a phase transition when the AdS soliton is chosen as the background  solution \cite{Surya:2001vj}. More recent results include the study of the thermodynamics of asymptotically AdS black holes of non-spherical topology in higher derivative gravity \cite{Nojiri:2002qn}, Gauss-Bonnet and dilaton gravity \cite{Cai:2007wz}, gravity dual to conformal field theory on a $S^2\times S^1\times \mathbb R$ boundary \cite{Copsey:2006br},
charged Ricci flat AdS black holes \cite{Banerjee:2007by} and R-charged black holes
\cite{Cai:2007vv}. Hence, the thermodynamics of black holes of nonspherical topology on the brane is well understood. It is therefore natural to see if the connection of brane and bulk thermodynamics persists for such AdS black holes. We will do so in this paper for a key case;  we will examine the connection between the thermodynamics of AdS black holes with Ricci flat horizons on the brane and their bulk description.  This case is especially interesting due to certain novel features of the thermodynamics of flat AdS black holes.

Recall that the thermodynamics of Schwarzschild-AdS black holes is determined by the black hole temperature $T$ and the AdS curvature radius $l$. Unlike the asymptotically flat case, the mass and horizon area of the Schwarzschild-AdS black hole are not uniquely specified by  temperature; generically there is both a large and small black hole for a given temperature. The large black hole has positive specific heat and the small one has negative specific heat. The free energy of the 4-dimensional black hole relative to that of thermal AdS is negative for $T> \frac {1}{\pi l}$ and positive for $\frac{\sqrt{3}}{2\pi l}<T< \frac {1}{\pi l}$. Hence the black hole is stable for $T> \frac {1}{\pi l}$ and is locally stable with tunnelling possible between it and thermal AdS for $\frac{\sqrt{3}}{2\pi l}<T< \frac {1}{\pi l}$. For $T<\frac{\sqrt{3}}{2\pi l}$, the black hole has negative specific heat and the only stable state is  thermal AdS spacetime.

Flat AdS black holes, locally asymptotically AdS spacetimes with Ricci flat horizons, exhibit quantitatively different behavior. Ricci flat AdS black holes also exhibit a phase transition when the AdS soliton is chosen as the background  solution \cite{Surya:2001vj}. However, unlike the Schwarzschild-AdS case, the stability of these two phases is determined by two variables, the temperature and  horizon circumference; hot, very small AdS solitons are stable as are  cold, very large flat AdS black holes. Furthermore,  the curvature radius  $l$ does not characterize the temperature of the phase transition. Additionally, the specific heat of the flat AdS black hole is always positive. Hence, in contrast to the Schwarzschild-AdS case,  flat AdS black holes are thermodynamically stable for all temperatures.

Given these important differences in the thermodynamic behavior, it is natural to investigate  whether or not they persist in the thermodynamics of  a braneworld scenario in which the KR branes contain flat AdS black holes.  We do so in Section \ref{thermo}.
We find that the bulk thermodynamics again precisely parallels that of the braneworld. Due to the nature of the phase transition in flat AdS black hole and black string spacetimes, a direct comparison of the phase transition temperature of a 5-dimensional flat AdS black hole to that of a flat AdS black string cannot be done. However, we can compute the possibility of  a phase transition from a flat AdS black string to a 5-dimensional AdS soliton and show that, though it can occur for certain thin black strings, the transition to the AdS soliton string is preferred.  In contrast to the Schwarzschild-AdS string case, the specific heat of the flat AdS black string is always positive. Hence we find that that the bulk solution consisting of a flat AdS black string is thermodynamically stable. It is consequently interesting to examine the perturbative stability of the flat AdS black string as the generalization of the Gubser-Mitra conjecture \cite{Gubser:2000mm} would imply perturbative stability for this case. We carry out a numerical analysis of this in Section \ref{pert}. We find that, in contrast to the Schwarzschild-AdS case, flat AdS black strings are always perturbatively stable. Hence, although the flat AdS black string does not exhibit translational invariance, the Gubser-Mitra conjecture holds.

\section{The thermodynamics of flat asymptotically AdS string spacetimes} \label{thermo}

Locally asymptotically AdS spacetimes in five dimensions with negative cosmological constant foliated by hypersurfaces of negative constant  Ricci curvature can be written as
\begin{align}
ds^2&=H^{-2}(z)\left(\frac {1}{l_4^2}\hat{g}_{\mu\nu}dx^\mu dx^\nu+dz^2\right)  \label{family}\\
H(z)&=\frac{\sin{z}}{l} \label{Hdef}
\end{align}
where
$l$ is the curvature radius of the 5-dimensional locally asymptotically AdS spacetime and $l_4$ that of the 4-dimensional one. These curvature radii are related to their respective cosmological constants by $\Lambda_5=-6/l^2$ and $\Lambda_4=-3/l_4^2$.

The metric (\ref{family}) is the bulk solution in a KR braneworld model \cite{Karch:2000ct} in which the brane tension is slightly detuned from its critical Randall-Sundrum value.
KR branes correspond to boundaries of this bulk solution at constant $z$.  The most general case is that of
two branes,  at $z_0= \pi/2 +\zeta_1$ and $z_1= \pi/2 -\zeta_0$, both with positive brane tension. These branes excise the  boundary at infinity of the locally asymptotically AdS spacetime; hence each provides a UV cutoff CFT that communicate via transparent boundary conditions.  The physical curvature radius of a 4-dimensional KR brane is set by its location; for a brane  at $\pi/2+\zeta_i$, $l_i=\frac{ l}{\cos \zeta_i}$.
 The special case $\zeta_1=\zeta_0 = \zeta$
 is of particular interest.  If the orbifold projection $z \to -z$ is also imposed, it yields a single AdS brane  with brane tension $\frac {6\sin \zeta}{8\pi G}$ and curvature radius $\frac{ l}{\cos \zeta}$  whose holographic dual has reflecting boundary conditions.\footnote{ Note that reflecting boundary conditions are those imposed in the usual gravitational analysis of Schwarzschild-AdS black holes. }

Two well-known solutions whose conformal boundary at infinity admits a Ricci flat metric are the flat AdS black hole \cite{Vanzo:1997gw} and the AdS soliton \cite{Horowitz:1998ha}. The AdS soliton metric can be written as
\begin{align}
\hat{g}^\ss_{\mu\nu}dx^\mu dx^\nu&=-r^2dt^2+\frac{dr^2}{V_{\ss}(r)}+V_{\ss}(r)d\theta^2+r^2d\phi^2\label{ts}\\
V_{\ss}(r)&=\frac{r^2}{l_4^2}-\frac{k^3_{\ss}}{r}\nonumber
\end{align}
where $\Lambda_4=-3/l_4^2$.
This spacetime is regular everywhere if $r\ge r_{\ss+}$ where $V(r_{\ss+})=0$ (hence $r_{\ss+}=k_\ss l_4^{2/3}$) and $\theta$ is periodic with period
\begin{equation*}
\alpha_{\ss}=4\pi\left(\left.\frac{\partial V_{\ss}(r)}{\partial r}\right|_{r=r_{\ss+}}\right)^{-1}=\frac{4\pi l_4^2}{3r_{\ss+}} \ .
\end{equation*}
The topology of its conformal boundary at infinity is  $S^1\times { \mathbb R}^2$
or $T^2\times {\mathbb R}$ if $\phi$ is  identified with period $2\pi$.

The flat AdS black hole metric can be written as\footnote{This is a special case of the more general metric of \cite{Vanzo:1997gw}. We choose this form for clarity; however our results are easily extended to the general case. This extension does not change our conclusions.}
\begin{align}
\hat{g}^\bs_{\mu\nu}dx^\mu dx^\nu&=-V_{\bs}(r)dt^2+\frac{dr^2}{V_{\bs}(r)}+r^2d\theta^2+r^2d\phi^2\label{tbh}\\
V_{\bs}(r)&=\frac{r^2}{l_4^2}-\frac{k^3_{\bs}}{r}\nonumber
\end{align}
The horizon, at $V_\bs(r_{\bs+})=0$ ($r_{\bs+} =k_\bs l_4^{2/3}$),  has topology  $S^1\times { \mathbb R}^2$ if
$\theta$ is identified with period $\alpha_\bs$
or $T^2\times {\mathbb R}$ if $\phi$ is also periodically identified. The conformal boundary at infinity has the same topology as the horizon and thus the same topology as that of the AdS soliton.

The AdS soliton string,  the 5-dimensional string solution (\ref{family}) constructed with the AdS soliton (\ref{ts}), is a regular, asymptotically AdS spacetime  with topology  $D^2\times  {\mathbb R}^3$ or $D^2\times S^1\times  {\mathbb R}^2$ in the respective cases.  The flat AdS black string, that constructed with the flat AdS black hole (\ref{tbh}),   has a spacetime topology exterior to the horizon of $S^1 \times {\mathbb R}^4$ or $T^2\times {\mathbb R}^3$ respectively. Both solutions  are foliated at large $r$ by a family of timelike surfaces with Ricci flat spatial slices of the same topology.

The Wick rotation $t\to i\tau$ of the AdS soliton string results in the Euclidean metric
\begin{equation}
ds_{\ss}^2=\frac{l^2}{l_4^2\sin^2 z}\left(r^2 d\tau^2+\frac{dr^2}{V_{\ss}(r)}+V_{\ss}(r)d\theta^2+r^2d\phi^2+l_4^2dz^2\right) \ .  \label{ess}
\end{equation}
Note that regularity places no constraint on $\tau$ for this metric; it can be chosen to have any (or no) periodicity.
The Wick rotation $t \to i \tau_b$ of the flat AdS black string results in
\begin{equation}
ds_{\bs}^2=\frac{l^2}{l_4^2\sin^2 z}\left(V_{\bs}(r) d\tau^2+\frac{dr^2}{V_{\bs}(r)}+r^2 d\theta^2+r^2d\phi^2+l_4^2dz^2\right)  \label{ebhs}
\end{equation}
where regularity now requires that $\tau_b$ be identified with period
\begin{equation*}
\beta_{\bs}=4\pi\left(\left.\frac{\partial V_{\bs}(r)}{\partial r}\right|_{r=r_{\bs+}}\right)^{-1}=\frac{4\pi l_4^2}{3r_{\bs+}}\ .\end{equation*}
A second Wick rotation of $\theta \to -it$ and a relabelling of coordinates $\tau_b \to \theta_s$ results in the AdS soliton string. Thus the same Euclidean instanton results from Wick rotation of both the AdS soliton string and the flat AdS black string.

The free energy of the flat AdS black string relative to that of the AdS soliton string determines the possibility of a phase transition between these two spacetimes. The difference in the free energy is proportional to the difference of their Euclidean actions. The Euclidean action for Einstein gravity with cosmological constant $\Lambda$ in $n$ dimensions is
\begin{equation*}
I=-\frac{1}{16\pi G}\int_{M} d^nx\sqrt{g}(R-2\Lambda)-\frac{1}{8\pi
G}\int_{\partial M} d^{n-1}x\sqrt{h}K
\end{equation*}
where $h$ is the induced metric, $K$ the extrinsic curvature of the boundary $\partial M$ and $G$ is the $n$-dimensional gravitational constant.
This action diverges for asymptotically AdS solutions; therefore it must be computed using a regularization procedure. We will do so by calculating the action at a finite radius $R$, find the difference
in the actions as a function of $R$ and then take the limit as $R\to \infty$.
Key to this procedure is to
match the induced geometry on the hypersurface at $R$ of the two solutions.
This matching requires that the periodicities of the coordinates for each solution are related by
\begin{equation} \beta_{\bs}\sqrt{V_{\bs}(R)} = R\beta_{\ss}\ \ \ \ \ \  R\alpha_{\bs}=\alpha_{\ss} \sqrt{V_{\ss}(R)} \ \ \ \ \ \  \gamma_{\bs}=\gamma_{\ss} = \gamma \ . \label{relations}
\end{equation}
For the case at hand, the difference of boundary terms  at fixed $R$ vanishes as $R\to\infty$.\footnote{The  corresponding extrinsic curvature boundary term for the KR brane is cancelled by the contribution to the free energy from the brane tension.}  Thus the action of the flat AdS black string relative to that of the AdS soliton string is simply
\begin{equation*}
I_{\bs}-I_{\ss}=\frac{l^3}{2\pi G l^4_4}\int^{z_1}_{z_0} \frac{dz} {\sin^5 z}\left( \beta_{\bs}\alpha_{\bs}\gamma\int_{r_{\bs+}}^Rr^2dr-\beta_{\ss}\alpha_{\ss}\gamma\int_{r_{\ss+}}^Rr^2dr\right)
\end{equation*}
which, after substitution from (\ref{relations}) and expansion in $R$, yields in the limit that $R\to \infty$, yields
\begin{align}
\Delta I &=
\frac{16 \pi^2l^3l_4}{81G }\left(\frac{\gamma\alpha_\ss}{\beta_\bs^2}\right)\left(\frac{\beta_\bs^3}{\alpha^3_\ss} - 1\right)\int^{z_1}_{z_0} \frac{dz}{ \sin^5 z}\nonumber\\
&= \frac{4G_4 l^3}{3 G l^2_4} \Delta I_4\left(C(\zeta_1) + C(\zeta_0)\right)\label{action}\\
C(u) &=\frac 14\frac {\sin u}{\cos^4u} + \frac 38 \frac {\sin u}{\cos^2 u}
+\frac 38 \ln\left (\frac {1+\sin u}{\cos u}\right) \nonumber\end{align}
where $\Delta I_4$ is the action of the 4-dimensional flat AdS black hole relative to the 4-dimensional AdS soliton with gravitational constant $G_4$.  $C(\zeta_1) + C(\zeta_0)$ is the contribution from the bulk integral over $z$; it is positive and diverges as $\zeta_1 \to \pi/2$ or $\zeta_0 \to \pi/2$. Thus, as in the spherical case, the physics of the bulk phase transition is completely determined by that  on the brane.
The sign of the relative action can be either positive or negative, depending on the value of $\beta_\bs$ and $\alpha_\ss$. If $\beta_\bs > \alpha_\ss$ the flat AdS black string is unstable relative to the AdS soliton string. If  $\beta_\bs <\alpha_\ss$, the opposite situation holds.  At the Hawking-Page phase transition point, $\beta_\bs = \alpha_\ss$, or equivalently $k_\bs=k_\ss$. Hence, in contrast to the Schwarzschild-AdS case, the phase transition temperature is independent of the curvature scale.

To further clarify the properties of the phase transition, it is useful to recall the form of the area of a spatial cross section of the black hole horizon of (\ref{tbh}) with $T^2$ topology;
\begin{equation*}
A = \frac {\alpha_\bs \gamma}{\beta_\bs^2}\left(\frac{4\pi l^2_4}{3}\right)^2=\frac {\alpha_\ss \gamma}{\beta_\bs^2}\left(\frac{16\pi^2 l^3_4}{9}\right) \ .
\end{equation*}
Unlike Schwarzschild-AdS black holes, the area of the horizon is not a function only of temperature; it  also depends on two independent parameters $\alpha_\ss$ and $\gamma$, the circumferences, characterizing the periodic identification of the two flat coordinates.
The instability condition $\beta_\bs > \alpha_\ss$ corresponds to an inverse temperature $\beta_\bs$ larger than that of
a preferred horizon circumference $\alpha_\bs=\alpha_\ss/l_4$. However, note that  the two black hole circumferences are in fact physically equivalent. Either could be chosen to match with the AdS soliton by the physics of the phase transition. This means that the phase transition
occurs when  $\beta_\bs$ becomes larger than the smallest of the two
circumferences. Hence $\alpha_{\bs}$ is chosen to be the smallest of the two circumferences. Therefore black holes with one small horizon circumference become unstable to decay to an AdS soliton as their temperature is decreased. In contrast, they become stable as their temperature is increased.

The thermodynamic stability of the flat AdS black string and AdS soliton string exactly parallels that of this phase transition on the brane. The bulk phase transition between the flat AdS black string and AdS soliton string occurs precisely when the phase transition between the black hole and AdS soliton occurs on every AdS slice.  The temperature of the black hole on each slice varies; for fixed $z$, the flat AdS black hole metric (\ref{tbh}) is scaled by a constant $c =  \frac{l}{l_4\sin z}$. Under this constant scaling, the physical parameters of the metric scale as $\bar \beta_{\bs} = c\beta_\bs$, $\bar r_{\bs+} = cr_{\bs+}$ and  $\bar l = cl_4$. The scale constant $c$ has its minimum value at $z=\pi/2$; for this slice, the black hole has its maximum temperature, minimum cross-sectional area of the horizon and  intrinsic curvature radius of $l$, that of the 5-dimensional spacetime. As $c$ increases, the temperature decreases and the cross-sectional area of the horizon increases; in fact, the decrease in temperature  for slices away from $z= \pi/2$ is exactly matched by the increase in black hole size. Therefore the flat AdS black hole on each constant $z$ slice is in equilibrium with its thermal bath. In summary, black holes get big and cold as $z$ approaches the conformal boundaries of the locally AdS spacetime at $z=0$ and $z=\pi$.

In the limit that both circumferences go to infinity, the phase transition flat AdS black string to the AdS soliton string disappears. This limit is equivalent to taking the AdS soliton mass parameter $k_\ss$ to zero;  in this limit, the AdS soliton string metric becomes singular. Its geometry is also the same as that of the zero mass flat AdS black string. It is known that no phase transition occurs when the zero mass flat AdS black hole is taken as the background \cite{Vanzo:1997gw,Birmingham:1998nr}; hence this limit of the AdS soliton string reproduces the expected result.

The energy of the flat AdS black string can also be computed; using the cutoff method \cite{Horowitz:1998ha}, the energy is given by
\begin{equation*}E = -\frac 1{8\pi G} \int_S \sqrt{\gamma}d^3x N(K-K_0)\end{equation*}
where $K$ is the extrinsic curvature of a surface $S$ given by the intersection of a constant radius and constant time surface and $K_0$ is that of the surface of the same geometry embedded in a static asymptotically AdS spacetime. This yields,  for an AdS soliton string reference spacetime,
\begin{align}E_{\bs}-E_{\ss} &= \frac  {4\pi^2 l^3l_4}{27 G} \alpha_ss\gamma(
\frac 2{\beta_\bs^3}+\frac 1 {\alpha_\ss^3}) \int_{z_0}^{z_1}\frac {dz}{\sin^3(z)}\label{senergy}\nonumber\\
&= \frac{G_4 l^3}{G l^2_4} \Delta E_4 \left( \bar C(\zeta_1)+\bar C(\zeta_0)\right)\\
\bar C(u) &=\frac 12 \frac {\sin u}{\cos^2 u}
+\frac 12 \ln\left (\frac {1+\sin u}{\cos u}\right) \nonumber\end{align}
As expected, the black string energy increases with increasing brane separation and the linear energy density is a minimum at $z=\pi/2$.  The energy  can also be calculated using the respective zero mass backgrounds as a reference:
\begin{equation*}E_{\bs} = \frac  {8\pi^2 l^3l_4}{27 G} \frac {\alpha_\ss\gamma}
{\beta_\bs^3}\left( \bar C(\zeta_1)+\bar C(\zeta_0)\right) \ \ \ \ \  E_{\ss} = - \frac  {4\pi^2 l^3l_4}{27 G} \frac {\gamma}
{\alpha_\ss^2}\left( \bar C(\zeta_1)+\bar C(\zeta_0)\right)  \ . \end{equation*}
Their difference yields (\ref{senergy}). The entropy of the black string is also easily found to be
\begin{equation}S  = \frac {A}{4G} \int_{z_0}^{z_1}\frac {dz}{\sin^3(z)}=\frac {A}{4G} \left( \bar C(\zeta_1)+\bar C(\zeta_0)\right) \label{entropy}\end{equation}
where $A/4G$ is the 4-dimensional flat AdS black hole entropy.

It is easy to verify that on each  constant $z$ slice, the entropy, energy and free energy obey the thermodynamic relation $\beta F_4 = \Delta I_4 = \beta \Delta E_4 - S_4$. However, this is not the case
for the 5-dimensional quantities;  the integrands of (\ref{action}), (\ref{senergy}),  and (\ref{entropy}) have different $z$ dependence. This is because these quantities, as well as the inverse temperature $\beta$, vary in $z$ for the flat AdS black string; it is in local thermodynamic equilibrium. Thus the flat AdS black string behaves more like a star than like a black hole. Hence (\ref{action}), (\ref{senergy}) and (\ref{entropy}) are properly viewed as the total value of the corresponding thermodynamic quantities. Furthermore, there will be no natural global definition of a nontrivial temperature for this configuration.\footnote{Of course, this flat AdS black string spacetime has zero temperature at infinity, in common with all other locally asymptotically AdS spacetimes.} Note this effect is not due to the choice of regularization method.  Other regularization methods, such as the counterterm method \cite{Emparan:1999pm},  may yield different numerical factors between terms in the computation of the relative action and energy.
However, use of a different regularization method will not change the differences in scaling of the $z$ integrals seen here.

Given  the proportionality of (\ref{senergy}) to $\Delta E_4$,  it is apparent that specific heat of the black string, $C = \frac {dE}{dT}$ is  positive and proportional to the integrated cross-sectional area of the black hole on each AdS slice. This result is, as one would anticipate, entirely due to the behavior of the specific heat for the flat AdS black hole itself.  This thermodynamic stability is associated with the nontrivial topology of these spacetimes; it may indicate an obstruction to the evaporation of the flat AdS black string due to the nontrivial topology of its spatial cross sections.

It is also interesting to consider whether or not there are other asymptotically AdS solutions
with the same topology of the boundary at infinity that contribute to the thermodynamics of the KR branes. Two natural candidates that can be chosen to have boundary topology $T^2 \times {\mathbb R}^2$ or $S^1\times {\mathbb R}^3$ are the 5-dimensional flat AdS black hole
\begin{align}
ds_{\b}^2&=-f_\b(r')dt^2+f_\b^{-1}(r')dr'^2+r'^2d\theta^2+r'^2d\phi'^2+r'^2dz'^2\label{5dbh}\\
f_\b(r')&=\frac{r'^2}{l^2}-\frac{{k}_{\b}^4}{r'^2},\nonumber
\end{align}
and the 5-dimensional AdS soliton
\begin{align}
ds_\s^2&=-r'^2dt^2+f_\s^{-1}(r')dr'^2+f_\s(r')d\theta^2+r'^2d\phi'^2+r'^2dz'^2\label{5sol}\\
f_\s(r')&=\frac{r'^2}{l^2}-\frac{{k}_{\s}^4}{r'^2},\nonumber
\end{align}
where, as for the 4-dimensional case, regularity requires $\theta$ be identified with period \begin{equation*}\alpha_\s=\frac{\pi l^2}{r'_{\s+}}\end{equation*}
where $r'_{\s+} = k_\s l^{\frac 12}$. Regularity of the Euclidean instanton corresponding to (\ref{5dbh}) yields the inverse temperature
\begin{equation*}\beta_{\b} =\frac{\pi l^2}{r'_{\b+}}\end{equation*}
where $r'_{\b+} = k_\b l^{\frac 12}$.
 The phase transition temperature of the 5-dimensional flat AdS black hole to the AdS soliton at $\beta_\b = \alpha_\s$ or equivalently $k_\b=k_\s$ is again independent of the curvature scale. Hence a comparison of the phase transition temperature of the black string to that of the 5-dimensional flat AdS black hole is not as straightforward as in the spherical case.

One can, however, compute the likelihood of a phase transition by comparing the action of the 5-dimensional flat AdS black hole relative to that of the zero mass AdS soliton to the action  of the flat AdS black string relative to the zero mass  AdS soliton string. This is possible as the
zero mass AdS soliton string (\ref{family}) and the zero mass 5-dimensional AdS soliton are locally the same spacetime; the coordinate transformation
\begin{equation*} r' = \frac {l r} {l_4\sin z} \ \ \ \ \ \   z'= \frac {l_4\cos z}{r}\ \ \ \ \ \  \theta' =\frac l{l_4} \theta
\ \ \ \ \ \  t'=t \ \ \ \ \ \  \phi' = \phi \label{coodtrans}\end{equation*}
brings the zero mass AdS soliton in the form (\ref{5sol}) to the metric of a zero mass soliton string,  (\ref{family}) with (\ref{ts}) as the 4-dimensional metric.\footnote{A similar coordinate transformation relates the zero mass  5-dimensional flat AdS black hole to the zero mass flat AdS black string. In fact, the zero mass AdS soliton and zero mass black hole are locally the same solution.} A KR brane boundary at $z_1$ lies on the curve $ z'r' = l\cot z_1$. Hence the zero mass AdS soliton string is a coordinatization of  the 5-dimensional zero mass AdS soliton whose brane boundaries intersect the boundary at infinity of the 5-dimensional AdS soliton at a cusp. The zero mass solutions are singular; hence any periodic identification of the angular coordinate $\theta'$ is allowed. Therefore, in contrast to earlier calculations,  the periodicity of  the angular coordinates are now fixed entirely in terms of flat AdS black string parameters;
\begin{equation}
\beta'_{0} = \beta_0 = \frac{\beta_\bs}{l_4}\ \ \ \ \ \ \ \alpha'_{0}=\frac{l}{l_4}\alpha_0=l\alpha_\bs\ \ \ \ \ \ \ \gamma'_{0}=\gamma_0=\gamma\label{solitonmatching}
\end{equation}
that is, the periodicities $\beta'_{0}, \alpha'_0, \gamma'_0$ of the zero mass 5-dimensional  AdS soliton are set by those of the zero mass AdS soliton string, $\beta_{0}, \alpha_0, \gamma_0$,  which, in turn, are set by those of the flat AdS black string.

The action of the flat AdS black string relative to the zero mass AdS soliton string is easily seen to be
\begin{equation}
\Delta I_{\bs} =-
\frac{16 \pi^2l^3 l_4^2}{81G}\left(\frac{\gamma\alpha_\bs}{\beta_\bs^2}\right)\left(C(\zeta_1) + C(\zeta_0)\right) \ . \label{bsactionrel0}\end{equation}
Similarly, the action of the AdS soliton string relative to the zero mass one is
\begin{equation}
\Delta I_{\ss} =- \frac{16 \pi^2l^3 l_4^2}{81G}\left(\frac{\beta_\ss\gamma}{\alpha_\ss^2}\right)\left(C(\zeta_1) + C(\zeta_0)\right) \ .\label{ssactionrel0}\end{equation}

That of the 5-dimensional black hole relative to the zero mass AdS soliton is given by
\begin{equation*}
I_{5\b}-I_{0}=\frac{1}{2\pi G l^2}\left( \beta_{\b}\alpha_{\b}\gamma\int_{r'_{\b+}}^Rr'^3dr'-\beta'_{0}\alpha'_{0}\gamma \int_{0}^Rr'^3dr'\right)\int^{\frac{l\, \cot z_1}{r'}}_{\frac{l\, \cot z_0}{r'}} dz' \ .
\end{equation*}
Note that the brane positions in the 5-dimensional flat AdS black hole has been set by their positions in the zero mass AdS soliton solution. After evaluation using the matching conditions
 $\beta'_{0}=\sqrt{f_\b(R)}\beta_\b/R $, $\alpha'_{0} = l\alpha_\b$ and taking the limit $R\to\infty$, one finds
\begin{align}\label{bactionrel0}
\Delta I_{5\b}&=-\frac{1}{6\pi G l}\left(\frac {\sin\zeta_1}{\cos\zeta_1}+ \frac {\sin\zeta_0}{\cos\zeta_0} \right) \beta_{\b}\alpha_{\b}\gamma r'^3_{\b+}\nonumber \\
&= - \frac{\pi^2l^5}{6G}\frac{\alpha_\b\gamma}{\beta^2_\b}\left(\frac {\sin\zeta_1}{\cos\zeta_1}+ \frac {\sin\zeta_0}{\cos\zeta_0} \right)
\end{align}

The action of the 5-dimensional AdS  soliton relative to the zero mass one is easily found by a similar computation:

\begin{align}\label{sactionrel0}
\Delta I_{5\s}&= - \frac{\pi^2l^5}{6G }\frac{\beta_\s\gamma_\s}{\alpha^2_\s}\left(\frac {\sin\zeta_1}{\cos\zeta_1}+ \frac {\sin\zeta_0}{\cos\zeta_0} \right) \ .
\end{align}

The differences in relative action can now be taken for all  four possible cases. First, note
that the matching (\ref{solitonmatching}) determines the 5-dimensional black hole temperature in terms of the flat AdS black string;
$\beta_\b = \frac{l}{l_4}\beta_\bs$, $\alpha_\b=\alpha_\bs$. The action for the flat AdS black string relative to the 5-dimensional black hole is then
\begin{equation}
\Delta I_{\bs}-\Delta I_{5\b}=  - \frac{\pi^2l^3l_4^2}{162G}\frac{\alpha_\bs\gamma}{\beta_\bs^2}\left (32C(\zeta_1) -
27\frac {\sin\zeta_1}{\cos\zeta_1}+32C(\zeta_0)-
27\frac {\sin\zeta_0}{\cos\zeta_0}\right) \ .
\end{equation}
Although this difference depends on the brane positions, it is easy to see that the overall sign is independent of this choice; $C(\zeta_i)$ is a monotonically increasing function that goes like $\sin\zeta_i$ for $\zeta_i$ sufficiently near zero. Therefore, the flat AdS black string with horizon radius $r_{\bs+}$ is stable relative to the 5-dimensional flat AdS black hole of radius $r'_{\b+}=\frac{l }{4l_4}r_{\bs+}$.

The action of the flat AdS black string relative to the 5-dimensional AdS soliton is, noting that (\ref{solitonmatching}) and the usual matching conditions give  $\beta_\s = \beta_\bs/l_4$, $\alpha_\s=l\alpha_\bs$,
\begin{equation}
\Delta I_{\bs}-\Delta I_{5\s}=  - \frac{\pi^2l^3l^2_4}{162G}\frac{\alpha_\bs\gamma}{\beta_\bs^2}\left (32C(\zeta_1) -
27\frac{\beta^3_\bs}{l^3_4\alpha^3_\bs}\frac {\sin\zeta_1}{\cos\zeta_1}+32C(\zeta_0)-
27\frac{\beta^3_\bs}{l^3_4\alpha^3_\bs}\frac {\sin\zeta_0}{\cos\zeta_0}\right) \ .
\end{equation}
In contrast to 5-dimensional flat AdS black hole case, the difference in actions can now have either sign as $\alpha_\bs$ is a free parameter. At small $\zeta_i$, one sees that $27\frac{\beta^3_\bs}{l^3_4\alpha^3_\bs}>32$, the flat AdS black string will be unstable to the 5-dimensional AdS soliton. As $\zeta_i$ becomes larger, this will no longer be the case; instead the transition depends explicitly on the brane position:
\begin{equation*}\frac{\beta^3_\bs}{l^3_4\alpha^3_\bs}>\frac{32(C(\zeta_1)+ C(\zeta_0))}{27(\frac {\sin\zeta_1}{\cos\zeta_1}+\frac {\sin\zeta_0}{\cos\zeta_0})}\ .\end{equation*}

As $C$ is increasing, the phase transition will occur only for smaller and smaller $\alpha_\bs$ as the branes move outward toward the boundary. In addition, recall that the transition between the flat AdS black string and AdS soliton string occurs when $\beta_\bs/\alpha_\ss=\beta_\bs/(l_4\alpha_\bs) =1$. Hence a flat AdS black string with a very small horizon circumference is also unstable to a phase transition to an AdS soliton string.  Therefore, a flat AdS black string of large horizon circumference will be stable and one with small horizon circumference will undergo a phase transition either to a 5-dimensional AdS soliton or an AdS soliton string.

It is easy to see that the action for the AdS soliton string relative to the 5-dimensional AdS soliton is
\begin{equation}
\Delta I_{\ss}-\Delta I_{5\s}=  - \frac{\pi^2l^3l_4^2}{162G}\frac{\beta_\ss\gamma}{\alpha_\ss^2}\left (32C(\zeta_1) -
27\frac {\sin\zeta_1}{\cos\zeta_1}+32C(\zeta_0)-
27\frac {\sin\zeta_0}{\cos\zeta_0}\right) \
\end{equation}
and that for the AdS soliton string relative to the 5-dimensional black hole is
\begin{equation}
\Delta I_{\ss}-\Delta I_{5\b}=  - \frac{\pi^2l^3l^2_4}{162G}\frac{\beta_\ss\gamma}{\alpha_\ss^2}\left (32C(\zeta_1) -
27\frac{\alpha_\ss^3}{l^3_4\beta^3_\ss}\frac {\sin\zeta_1}{\cos\zeta_1}+32C(\zeta_0)-
27\frac{\alpha_\ss^3}{l^3_4\beta^3_\ss}\frac {\sin\zeta_0}{\cos\zeta_0}\right) \ .
\end{equation}

From the above results, it is clear that the 5-dimensional AdS soliton is always unstable to the AdS soliton string. Thus if a  flat AdS black string with a very small horizon circumference undergoes a phase transition to the 5-dimensional AdS soliton, as this is only a local minimum it in turn will undergo a phase transition to the AdS soliton string.  Similarly, there is a phase transition between the 5-dimensional flat AdS black hole and the AdS soliton string with stability determined by the black  hole parameters.

At first glance, these results seem to contradict the conjecture that the AdS soliton is the ground state of an asymptotically AdS spacetime with Ricci flat boundary at infinity. However, this is not the case; the above results are comparing not a pure 5-dimensional AdS soliton, but rather one with brane boundaries. These branes in the 5-dimensional AdS soliton space are held in position by their tension; hence, one is not comparing a bulk 5-dimensional AdS soliton, but rather one constrained by these branes. Furthermore, the brane geometry induced by slicing the 5-dimensional AdS soliton
\begin{equation}
ds^2=-r'^2dt^2+\left(f_\s^{-1}(r')+l^2\frac{\cot^2z_1}{r'^2}\right)dr'^2+f_\s(r')d\theta'^2+r'^2d\phi^2
\end{equation}
is not particularly nice. Although the spacetime is asymptotically a 4-dimensional  locally AdS spacetime
with physical curvature scale $l_p = \frac l{\sin^2 z_1}$ as expected, the mass of the spacetime is zero. Its Ricci tensor obeys the null energy condition,\footnote{ The null energy condition is that
$R_{ab}k^ak^b \ge 0$ for all null vectors $k^a$.} but exhibits radial dependence and is anisotropic.
Hence  the tension of this brane will also exhibit a corresponding unusual form.

Of course these calculations are somewhat unsatisfactory. Although the same space is used in the calculation of the relative actions,  the cutoff at constant radius used in the calculation of (\ref{bsactionrel0}), (\ref{ssactionrel0}) and in  (\ref{bactionrel0}), (\ref{sactionrel0})  occurs on different surfaces for each case. However, if the limit is well defined, the interchange of the difference and limit should not affect the result.  Moreover, there is a certain physical appeal to these results that leads one to expect that they hold at least qualitatively;
although in certain circumstances, a flat AdS black string can be unstable to a 5-dimensional AdS soliton with an unusual brane geometry, any such configuration is also unstable to an AdS soliton string. Hence all paths lead to the same final outcome.

\section{ The perturbative stability of asymptotically  Ricci flat AdS  string spacetimes} \label{pert}

Unlike the Schwarzschild black hole which has been proven to be perturbatively stable \cite{Vishveshiwara}, it is well known that black strings and black p-branes can be perturbatively unstable \cite{Gregory:1993vy,Gregory:1994bj} ( See \cite{Harmark:2007md} for a recent review). This instability is due to the existence of a metric perturbation that diverges exponentially in time associated with a translational symmetry of the string spacetime.
The Gubser-Mitra conjecture \cite{Gubser:2000mm} connects this classical perturbative instability of the spacetime to its thermodynamic instability, namely a negative specific heat.  This conjecture has been proven  for certain classes black branes with a non-compact translational symmetry \cite{Reall:2001ag,Ross:2005vh}.

However, the case in which there is a non-compact  translational symmetry on a conformally related space is also relevant. The perturbative stability of black AdS strings with Schwarzschild spacetime cross sections was first analysed by Gregory in  \cite{Gregory:2000gf}.
Hirayama and Kang \cite{Hirayama:2001bi} then studied the perturbative stability of the black AdS string constructed from the 4-dimensional Schwarzschild-AdS metric.
They found numerically that here is a threshold value, proportional to the ratio of the 4-dimensional AdS curvature radius $l_4$ to the string mass parameter $k_\bs$, above which the black string is perturbatively stable. For values of the mass parameter smaller than this threshold value, the string is unstable. It is this instability that was associated with the negative specific heat of the black string solution in \cite{Chamblin:2004vr}. Hence, at least for the black AdS string constructed from Schwarzschild-AdS, the Gubser-Mitra conjecture holds with the weaker assumption of  a conformal  killing vector.

This generalization make sense for this case;
if we consider each AdS slicing as a spacetime in its own right, the Gubser-Mitra conjecture applies and identifies a perturbative instability associated with the temperature at which the specific heat becomes negative for the Schwarzschild-AdS. Scaling of temperature and horizon area indicates that this property persists for each slice of the black string.  Hence, though the space itself is not translationally invariant, its thermodynamics behaves in an invariant way with respect to the conformal killing vector $\partial_z$. In contrast,  the flat AdS black hole and the flat AdS black string have positive specific heat. Furthermore, the thermodynamics of the flat AdS black string is also invariant. Hence, we anticipate that the flat AdS black string will be perturbatively stable.

In this section we examine the perturbative stability of 5-dimensional AdS strings whose cross sections are AdS spacetimes with toroidal topology. Our analysis closely follows that of \cite{Hirayama:2001bi}. In the following we will only address the stability of the bulk AdS string solutions; we anticipate that the presence of branes, though changing certain details in the calculation, will not change our conclusions.

It is convenient to write the general tensor perturbation of the metric (\ref{family}) in the form
\begin{equation}
ds^2=H^{-2}(z)\left[(\hat{g}_{\mu\nu}+h_{\mu\nu}(x,z))dx^\mu
dx^\nu+dz^2\right]
\end{equation}
and impose transverse traceless gauge\footnote{ This gauge, is also termed Randall-Sundrum
gauge \cite{Randall:1999vf}.} on the brane
\begin{equation}\nabla^\mu h_{\mu\nu}(x)=0\ \ \ \ \ \ \ \
h_{\mu}^{\mu}(x)=0 \ .\label{gauge}\end{equation}

Let
$h_{\mu\nu}(x,z)=H^{3/2}(z)\xi(z)h_{\mu\nu}(x)$; then to linear order, the Einstein equations can be written
as two decoupled equations
\begin{align}
\nabla_\rho\nabla^\rho
h_{\mu\nu}(x)+2R_{\mu\rho\nu\tau}h^{\rho\tau}(x)&=m^2h_{\mu\nu}(x)\label{Lich}\\
[-\partial_z^2+W(z)]\xi(z)&=m^2\xi(z)\ \ \ \ \ \ W(z) =-\frac{3}{2}\frac{H''}{H}+\frac{15}{4}\left(\frac{H'}{H}\right)^2\label{zdirec}
\end{align}
where the covariant derivative, the Riemann tensor, and the raising and lowering of indices are with respect
to the 4-dimensional metric $\hat{g}_{\mu\nu}$.  Hence, as usual, the tensor perturbation (\ref{Lich}) takes the form of a massive graviton and the minimum mass of this perturbation is set by the minimum eigenvalue of (\ref{zdirec}).
Using (\ref{Hdef}), one finds that the explicit expression of the effective
potential is
$$W(z)=\frac{1}{l_4^2}\left(\frac{3}{2}+\frac{15}{4}\cot^2\frac{z}{l_4}\right)$$
which is bounded below by $\frac 32$ and becomes infinite at $z=n\pi l_4, n\in\mathbb{Z}$.
Thus (\ref{zdirec}) has the same spectrum as that of the Schwarzschild-AdS string  \cite{Hirayama:2001bi};  solving, one finds $m_{\textrm{min}}=2/l_4$. Therefore the difference between the spherical case and that of toroidal AdS slicings is due entirely to the different background geometry for the graviton in (\ref{Lich}).

For the toroidal AdS slicings, we shall seek a marginally unstable tensor perturbation exhibiting the symmetry of the spacetime;
\begin{equation}
h_{\mu\nu}=e^{\Omega t}\begin{pmatrix}
B_{tt}(r) & B_{tr}(r) & 0 & 0 \\
B_{tr}(r) & B_{rr}(r) & 0 & 0 \\
0 & 0 & B_{\theta\theta}(r) & 0 \\
0 & 0 & 0 & B_{\phi\phi}(r)
\end{pmatrix}\label{hmn} \ .
\end{equation}

At this point, the cases of the flat AdS black string and AdS soliton string must be analysed separately.

\subsection{The flat AdS black string}

For the flat AdS black hole metric (\ref{tbh}), the gauge conditions (\ref{gauge}) yield the relations
\begin{align}
-\frac{\Omega}{V_\bs}B_{tt}+V_\bs B_{tr}'+\left(V_\bs'+\frac{2V_\bs}{r}\right)B_{tr}&=0\nonumber\\
-\frac{\Omega}{V_\bs}B_{tr}+V_\bs B_{rr}'+\frac{2V_\bs}{r}B_{rr}+\frac{V_\bs'}{2V_\bs^2}B_{tt}+\frac{3V_\bs'}{2}B_{rr}-\frac{1}{r^3}(B_{\theta\theta}+B_{\phi\phi})&=0\nonumber\\
-\frac{1}{V_\bs}B_{tt}+V_\bs B_{rr}+\frac{1}{r^2}(B_{\theta\theta}+B_{\phi\phi})&=0 \ .\label{B}
\end{align}

Note $'$ has been used to indicate differentiation by $r$. The graviton
perturbation equations (\ref{Lich}) for the $tr$ and $rr$ components for the flat AdS black hole slicing are equivalent to
\begin{align}
r^2V_\bs^3B_{tr}''&+(r^2V_\bs'V_\bs^2+2rV_\bs^3)B_{tr}'\nonumber\\&+(-\Omega^2r^2V_\bs-2V_\bs^3
+r^2V_\bs''V_\bs^2-r^2V_\bs'^2V_\bs-m^2r^2V_\bs^2)B_{tr}\nonumber\\&+\Omega
r^2V_\bs'B_{tt}+\Omega r^2V_\bs'V_\bs B_{rr}=0\\
-2r^4V_\bs^4B_{rr}''&+(-6r^4V_\bs^3V_\bs'-4r^3V_\bs^4)B_{rr}'\nonumber\\
&+(2\Omega^2r^4V_\bs^2-2r^4V_\bs^3V_\bs''-r^4V_\bs^2V_\bs'^2-4r^3V_\bs^3V_\bs'+8r^2V_\bs^4+2m^2r^4V_\bs^3)B_{rr}\nonumber\\
&+(r^4V_\bs^2-2r^4V_\bs V_\bs'')B_{tt}-4\Omega r^4V_\bs V_\bs'B_{tr}+(-4V_\bs^3+2rV_\bs^2V_\bs')(B_{\theta\theta}+B_{\phi\phi})=0\label{E}
\end{align}

By eliminating $B_{tt}$, $B_{rr}$,
$B_{\theta\theta}$ and $B_{\phi\phi}$, the above equations reduce to
\begin{equation}
C_2B_{tr}''+C_1B_{tr}'+C_0B_{tr}=0\label{tr}
\end{equation}
where
\begin{align*}
C_2=&(8+4\mu^2)r^{14}+4\omega^2r^{12}+(-32-12\mu^2)r_{\bs+}^3r^{11}-8\omega^2r_{\bs+}^3r^9\\
&+(39+12\mu^2)r_{\bs+}^6r^8+4\omega^2r_{\bs+}^6r^6+(-14-4\mu^2)r_{\bs+}^9r^5-r_{\bs+}^{12}r^2\\
C_1=&(48+24\mu^2)r^{13}+32\omega^2r^{11}+(-168-48\mu^2)r_{\bs+}^3r^{10}-28\omega^2r_{\bs+}^3r^8\\
&+(108+24\mu^2)r_{\bs+}^6r^7-4\omega^2r_{\bs+}^6r^5+15r_{\bs+}^9r^4-3r_{\bs+}^{12}r\\
C_0=&(32+8\mu^2-4\mu^4)r^{12}+(24-8\mu^2)\omega^2r^{10}+(-176-12\mu^2+8\mu^4)r_{\bs+}^3r^9\\&-4\omega^4r^8
+(24+8\mu^2)\omega^2r_{\bs+}^3r^7+(12-9\mu^2-4\mu^4)r_{\bs+}^6r^6\\&-3\omega^2r_{\bs+}^6r^4+(52+13\mu^2)r_{\bs+}^9r^3-r_{\bs+}^{12}
\end{align*}
in which $\mu^2=m^2l_4^2$ is the dimensionless KK mass, $r_{\bs+}^3=k_\bs^3l_4^2$, the horizon radius
and $ \omega=\Omega l_4$.  Note that, in contrast to the Schwarzschild-AdS case, $r_{\bs+}$  is interchangeable with the mass parameter $k_\bs$; this is due to the form of $V_\bs$ for the flat AdS black hole. Furthermore, also due to the form of $V_\bs$, (\ref{tr}) is exactly invariant under the scaling $r\to \alpha r$, $r_{\bs+} \to \alpha r_{\bs+}$, $\omega\to \alpha \omega$. Thus by rescaling we can fix the value of either $r_{\bs+}$ or $\omega$; we will do so by taking $r_{\bs+} = 1$.
Now (\ref{tr}) only has two arbitrary constants, $\mu$ and $\omega$, in contrast to the Schwarzschild-AdS case \cite{Hirayama:2001bi}; this simplifies the analysis of the stability. We will solve (\ref{tr})
as an eigenvalue problem, treating
$\mu^2$ as a parameter and $\omega$ as the eigenvalue.

The asymptotic
behavior of solutions of (\ref{tr})  approaching the horizon and spatial infinity is
easily calculated;
\begin{align}
r&\rightarrow 1,\quad B_{tr}\sim (r-1)^{-1\pm\omega/3}\label{bchorizon}\\
r&\rightarrow\infty,\quad B_{tr}\sim
r^{(-5/2\pm\sqrt{9/4+\mu^2}\, )}\label{bcinfinity}
\end{align}

As discussed by \cite{ Gregory:1993vy,Gregory:1994bj}, the specification of the boundary conditions is key to a consistent perturbative solution. The perturbation must remain small exterior to the black hole horizon. As the negative root solution in (\ref{bchorizon}) diverges at the horizon, it should be excluded. Unfortunately, the positive root solution will also diverge as $r \to 1$ if $\omega < 3$. However, this divergence is attributable to the failure of the Schwarzschild-like coordinates used here. A change of coordinates to a set regular on the horizon, such as Kruskal coordinates, demonstrates that the perturbation will be regular if  it diverges more slowly than $(r-1)^{-1}$.  Therefore only the positive root of (\ref{bchorizon}) exhibits the correct behavior.

Furthermore,
an additional condition, not present in asymptotically flat spacetimes,  is necessary as AdS spacetimes are not globally hyperbolic. The usual condition for matter fields, and that imposed by us here, is that that the energy in the perturbation be finite \cite{Avis:1977yn,Breitenlohner:1982jf}. This requires a fall-off of $B_{tr} \sim r^\lambda $ with $\lambda < -5/2$ as $r\to\infty$. This is true of
the behavior of the negative root of (\ref{bcinfinity}), but not the positive one. Thus to summarize, the correct boundary conditions for the perturbation are $ B_{tr}\sim (r-1)^{-1+\omega/3}$ as $ r\to 1$
and $ B_{tr}\sim
r^{(-5/2-\sqrt{9/4+\mu^2}\, )} $ as $r\to \infty$.

To numerically solve (\ref{tr})  we transform it into the related nonlinear equation
by dividing it by $B_{tr}$. Then as $\frac{B_{tr}''}{B_{tr}}=\left(\frac{B_{tr}'}{B_{tr}}\right)'+\left(\frac{B_{tr}'}{B_{tr}}\right)^2$, (\ref{tr}) becomes
\begin{equation}
C_2(Y'+Y^2)+C_1Y+C_0=0\label{smeq}
\end{equation}
where $Y=B_{tr}'/B_{tr}$. This form is solvable by the shooting method. $Y$ will be regular everywhere if $B_{tr}$ is nonvanishing on the interior.
One anticipates that the marginally unstable mode will exhibit such behavior.
The boundary conditions can now be imposed in a straightforward manner;
$Y\sim (-1 + \omega/3)/(r-1) $ at the horizon and $Y \sim(-5/2-\sqrt{9/4+\mu^2}\, )/r$ as $r$ goes to infinity.
\footnote{The shooting method cannot be applied directly to (\ref{tr}) due to the
sensitivity of the asymptotic behavior to numerical errors. Numerical errors in the
integration could drive the solution to exhibit the behavior of the larger asymptotic
solution.  Then we would not be able to see the correct behavior by varying the
value of $\omega$.}

We used the shooting method algorithm of \cite{NRC} implemented in C. We tested this  code on
the equivalent equations for the asymptotically flat string of \cite{Gregory:1993vy} and the spherical AdS string of \cite{Hirayama:2001bi} and reproduced their results.
We then applied this code to (\ref{smeq}) and found no positive eigenvalue $\omega$ that solves the equation for any chosen value of the parameter $\mu^2$. To be explicit, starting from either boundary, and with any value of $\mu^2\ge4$, we find the value of $\omega$ rapidly diverges to infinity. This means that higher frequency modes solve (\ref{Lich}) and (\ref{zdirec}) better, which indicates that an unstable mode with finite energy does not exist. Therefore, numerical evidence indicates that the flat AdS black string is perturbatively stable.

\subsection{The AdS soliton string}

It is useful to compare the results of the flat AdS black string case to that of the AdS soliton string. The
steps are similar to those of the previous subsection.

The gauge conditions for the AdS soliton metric (\ref{ts}) are
\begin{equation}\begin{split}
-\frac{\Omega}{r^2}B_{tt}+V_\ss B_{tr}'+\left(V_\ss'+\frac{2V_\ss}{r}\right)B_{tr}&=0\label{Bs}\\
-\frac{\Omega}{r^2}B_{tr}+V_\ss B_{rr}'+\frac{2V_\ss}{r}B_{rr}+\frac{1}{r^3}B_{tt}+\frac{3V_\ss'}{2}B_{rr}-\frac{V_\ss'}{2V_\ss^2}B_{\theta\theta}-\frac{1}{r^3}B_{\phi\phi}&=0\\
-\frac{1}{r^2}B_{tt}+V_\ss B_{rr}+\frac{1}{V_\ss}B_{\theta\theta}+\frac{1}{r^2}B_{\phi\phi}&=0\ .
\end{split}\end{equation}
The graviton
perturbation equations (\ref{Lich}) for the $tr$ and $rr$ components are equivalent to
\begin{align}
2r^2V_\ss B_{tr}''&+(4rV_\ss+4r^2V_\ss')B_{tr}'\nonumber\\ &+(4rV_\ss'-2V_s+r^2V_\ss''-2\Omega^2-2m^2r^2)B_{tr}+4rV_\ss\Omega B_{rr}=0\\
(-4\Omega V_\ss^2&-\Omega r^2(V_\ss')^2+2\Omega rV_\ss V_\ss'+2\Omega r^2V_\ss V_\ss'')B_{tr}\nonumber\\
&+(r^4V_\ss^2V_\ss'-2r^3V_\ss^3)B_{rr}''+\left[-8r^2V_\ss^3-2r^3V_\ss'V_\ss^2-2r^4V_\ss^2V_\ss''+4r^4V_\ss(V_\ss')^2\right]B_{rr}'\nonumber\\
&+\big[2\Omega^2rV_\ss^2-\Omega^2r^2V_\ss V_\ss'+2r^4(V_\ss')^3-12r^2V_\ss^2V_\ss'+8r^3V_\ss(V_\ss')^2-8r^3V_\ss^2V_\ss''\nonumber\\
&-2r^4V_\ss V_\ss'V_\ss''+2m^2r^3V_\ss^2-m^2r^4V_\ss V_\ss'\big]B_{rr}=0
\end{align}

In contrast to the flat AdS black hole string, the AdS soliton equations are not invariant under the interchange
of $B_{\theta\theta}$ and $B_{\phi\phi}$. Nonetheless, everything can be expressed in terms of $B_{tr}$ albeit at a price; the equation is now fourth order,
\begin{equation}
C_4H_{tr}''''+C_3H_{tr}'''+C_2H_{tr}''+C_1H_{tr}'+C_0H_{tr}=0\label{fourth}
\end{equation}
where
\begin{align*}
C_4=&r^{10}-2r_{\ss+}^3r^7+r_{\ss+}^6r^4\\
C_3=&22r^9-26r_{\ss+}^3r^6+4r_{\ss+}^6r^3\\
C_2=&(142-2\mu^2)r^8-2\omega^2r^6+(-86+2\mu^2)r^5+2\omega^2r_{\ss+}^3r^3-2r_{\ss+}^6r^2\\
C_1=&(288-18\mu^2)r^7-14\omega^2r^5+(-72+6\mu^2)r_{\ss+}^3r^4+2\omega^2r_{\ss+}^3r^2\\
C_0=&(112-32\mu^2+\mu^4)r^6+(-10+2\mu^2)\omega^2r^4+(-8+2\mu^2)r_{\ss+}^3r^3\\&+\omega^4r^2-2\omega^2r_{\ss+}^3r+4r_{\ss+}^6\ .
\end{align*}

Again, the asymptotic behavior can be easily determined; as
$r$ approaches infinity
\begin{align}
B_{tr}&\sim
r^{(-11/2\pm\sqrt{9/4+\mu^2}\, )}\label{bcinfinity2}\\
B_{tr}&\sim
r^{(-5/2\pm\sqrt{9/4+\mu^2}\, )}\label{bcinfinity3}
\end{align}
For $B_{tr}$ to fall off faster than $r^{-5/2}$ as $r \to\infty$, the possible asymptotic behavior is $B_{tr}\sim C r^{(-5/2-\sqrt{9/4+\mu^2}\, )}+D r^{(-11/2-\sqrt{9/4+\mu^2}\, )}$ for $\mu^2>27/4$. The first term is dominant. If $\mu^2$ is smaller than this, the asymptotic behavior is $B_{tr}\sim C r^{(-5/2-\sqrt{9/4+\mu^2}\, )}+Dr^{(-11-\sqrt{9/4+\mu^2}\, )}+E r^{(-11/2+\sqrt{9/4+\mu^2}\, )}$  and the last term is dominant.

In contrast to the flat AdS black hole string case, the spacetime is regular as $r \to 1$. Therefore, the perturbation must be regular at this point. Thus $B_{tr}\sim (r-1)^{-1+\sqrt{3}}$ as $r\to 1$.

\begin{figure}
\centering
\includegraphics[width=0.4\textwidth]{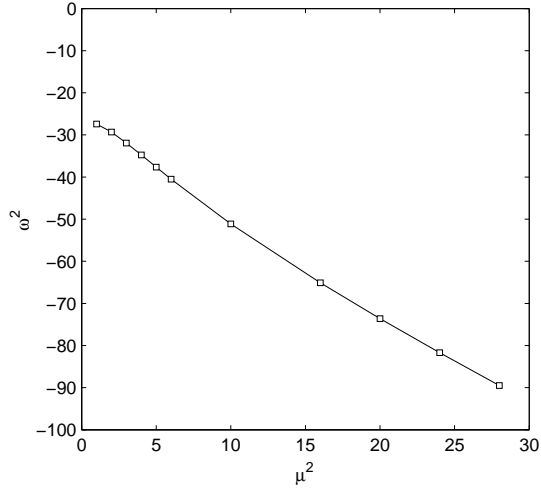}
\label{mOmega}
\caption{Dependence of first eigenvalue of $\omega^2$ on $\mu^2$. This graph shows that we find only negative eigenvalues of $\omega^2$, which correspond to stable oscillatory modes.}
\end{figure}

As in the flat AdS black string case, (\ref{fourth}) can also be rewritten in terms of $Y=B_{tr}'/B_{tr}$,
resulting in the third order equation
\begin{equation}
C_4(Y'''+4Y''Y+6Y'Y^2+3Y'^2+Y^4)+C_3(Y''+3Y'Y+Y^3)+C_2(Y'+Y^2)+C_1Y+C_0=0
\end{equation}
Again, no solution is found numerically using the shooting method code, both for the case where $\mu^2>27/4$
and for  $\mu^2<27/4$.
However, in contrast to the flat AdS black string case,  it is now possible to examine the behavior of this equation in more detail. The boundary conditions
are independent of $\omega$. Furthermore, (\ref{fourth}) depends only on $\omega^2$. These facts allow us to seek solutions with negative values of $\omega^2$. Of course, any such solution will correspond to a stable mode. In fact, we find such solutions; the numerical results are shown in Figure 1.
This numerical result confirms that the AdS soliton string is stable.
\section{Conclusions}

As in the Schwarzschild-AdS case \cite{Chamblin:2004vr}, we find that the thermodynamics of flat AdS black holes on KR branes is exactly paralleled by that of the flat AdS black string in the bulk. The flat AdS black string is in local thermodynamic equilibrium on each constant $z$ slice; thus its phase transition occurs in parallel with that of the flat AdS black hole transition on the brane. Notably, both the brane and bulk thermodynamics retain the unique characteristic of the flat AdS black hole that the phase transition temperature is independent of the curvature radius $l$.  Thus this model provides a forum for the study of the properties of black holes on asymptotically AdS branes and the AdS/CFT correspondence in which the phase transition temperature is independent of the curvature radius.

 In addition, we have shown, by a calculation of the relative action, that although very small circumference flat AdS black strings can be unstable to a 5-dimensional AdS soliton with an unusual brane geometry, any such configuration is also unstable to an AdS soliton string.
Furthermore, the flat AdS black string is stable relative to the 5-dimensional flat AdS black hole.  In contrast to the Schwarzschild-AdS case, the specific heat of the flat AdS black string is always positive. A numerical  calculation indicates that the bulk black AdS string and AdS soliton string are both perturbatively stable for all values of their mass parameter. Thus Gubser-Mitra conjecture holds for the flat AdS black string.

Although our analysis is confined to the 5-dimensional case, it is clear that the thermodynamics of higher dimensional flat AdS black strings will have qualitatively similar behavior. However,  it is not
obvious that the more quantitative features in the analysis will carry through as dimensionality is known to play an important role in asymptotically flat string perturbative stability and phase transitions \cite{Sorkin:2004qq,Kol:2004pn}.

It is also natural to consider whether or not these results will also hold true for AdS black strings whose spatial cross sections are hyperbolic AdS black holes. Hyperbolic AdS black holes also have manifestly positive specific heat \cite{Vanzo:1997gw,Brill:1997mf}. Hence one anticipates that hyperbolic AdS string spacetimes will exhibit perturbative stability. However, the thermodynamics of these spacetimes will be quite different in other aspects from that of the flat AdS black string  as there is no analog of the AdS soliton for  hyperbolic case \cite{Galloway:2001uv,Galloway:2002ai}. In particular, one expects that there will be no bulk phase transition for the hyperbolic AdS black string.

\section*{Acknowledgements}
This work was supported by the Natural Sciences and Engineering Council of Canada.
In addition, the KS and DW would like to
thank the Perimeter Institute for its hospitality during the completion of this paper.

\end{document}